\def \VersionLong {}
\ifdefined \VersionLong
	\newcommand{\LongVersion}[1]{\ifdefined\VersionWithComments{\color{red!40!black}#1}\else#1\fi}
	\newcommand{\ShortVersion}[1]{\ifdefined\VersionWithComments{\color{black!40}#1}\fi}
\else
	\newcommand{\LongVersion}[1]{\ifdefined\VersionWithComments{\color{black!40}#1}\fi}
	\newcommand{\ShortVersion}[1]{\ifdefined\VersionWithComments{\color{red!40!black}#1}\else#1\fi}
\fi

\documentclass[conference]{IEEEtran}
\IEEEoverridecommandlockouts
\usepackage{amsmath,amssymb,amsfonts}
\usepackage{graphicx}
\graphicspath{{figures/}}

\usepackage[utf8]{inputenc}
\usepackage[english]{babel}

\usepackage[ruled,vlined,linesnumbered]{algorithm2e}

\usepackage{subcaption}

\usepackage{paralist} %

\newenvironment{ienumeration}
	{\ifdefined\VersionLong\begin{enumerate}\else\begin{inparaenum}[\itshape i\upshape)]\fi}
	{\ifdefined\VersionLong\end{enumerate}\else\end{inparaenum}\fi}

\usepackage{amsmath} %
\usepackage{amssymb} %

\LongVersion{
	\usepackage[backend=biber,backref=true,style=alphabetic,url=false,doi=true,defernumbers=true,sorting=anyt,maxnames=99]{biblatex} %
	\addbibresource{Time4sys.bib}

	\renewbibmacro*{doi+eprint+url}{%
		\iftoggle{bbx:doi}
			{\color{black!40}\footnotesize\printfield{doi}}
			{}%
		\newunit\newblock
		\iftoggle{bbx:eprint}
			{\usebibmacro{eprint}}
			{}%
		\newunit\newblock
		\iftoggle{bbx:url}
			{\usebibmacro{url+urldate}}
			{}%
	}

}

\ifdefined\VersionWithComments
	\usepackage{draftwatermark}
	\SetWatermarkText{draft}
	\SetWatermarkScale{3}
	\SetWatermarkColor[gray]{0.9}
\fi
\usepackage[svgnames,table]{xcolor}
\definecolor{darkblue}{rgb}{0.0,0.0,0.6}
\definecolor{darkgreen}{rgb}{0, 0.5, 0}
\definecolor{darkpurple}{rgb}{0.7, 0, 0.7}
\definecolor{darkblue}{rgb}{0, 0, 0.7}

\LongVersion{
\usepackage[
		colorlinks=true,
	\ifdefined \VersionWithComments
		pagebackref=true,
	\fi
		citecolor=darkgreen,
		linkcolor=darkblue,
		urlcolor=darkpurple,
	]{hyperref}
}

\ShortVersion{
	\newcommand{\url}[1]{\texttt{#1}}
}
	
\usepackage[capitalise,english,nameinlink]{cleveref} %
\crefname{line}{\text{line}}{\text{lines}} %

\usepackage{tikz}
\usetikzlibrary{arrows,automata}
\tikzstyle{every node}=[initial text=]
\tikzstyle{location}=[rectangle, rounded corners, minimum size=12pt, draw=black, fill=blue!10, inner sep=2pt]
\tikzstyle{invariant}=[draw=black, dotted, inner sep=1pt] %
\tikzstyle{urgent}=[dotted, draw=red, very thick]

\tikzstyle{activation} = [thick,->]
\tikzstyle{deadline} = [very thick,->]
\tikzstyle{violation} = [draw=red,very thick]

\definecolor{loccolor1}{rgb}{.9, .95, 1}
\definecolor{loccolor2}{rgb}{.9, .95, 1}
\definecolor{loccolor3}{rgb}{.9, .95, 1}
\definecolor{loccolor4}{rgb}{.9, .95, 1}
\definecolor{loccolor5}{rgb}{1, .5, .5} %

\definecolor{coloract}{rgb}{0.50, 0.70, 0.30}
\definecolor{colorclock}{rgb}{0.4, 0.4, 1}
\definecolor{colordisc}{rgb}{1, 0, 1}
\definecolor{colorloc}{rgb}{0.2, 0.2, 0.35}
\definecolor{colorparam}{rgb}{1, 0.6, 0.0}

\newcommand{\styleact}[1]{\ensuremath{\textcolor{coloract}{\mathrm{#1}}}}
\newcommand{\styleclock}[1]{\ensuremath{\textcolor{colorclock}{\mathrm{#1}}}}

\newcommand{\styleloc}[1]{\ensuremath{\textcolor{colorloc}{\mathrm{#1}}}}
\newcommand{\styleparam}[1]{\ensuremath{\textcolor{colorparam}{\mathrm{#1}}}}

\newcommand{\init}{_0}

\newcommand{\A}{\ensuremath{\mathcal{A}}}
\newcommand{\Actions}{\Sigma}
\newcommand{\action}{\ensuremath{a}}

\newcommand{\C}{C}

\newcommand{\Clock}{\mathbb{X}} %
\newcommand{\ClockCard}{H} %
\newcommand{\clock}{x} %
\newcommand{\clockval}{\nu} %
\newcommand{\ClocksZero}{\vec{0}}
\newcommand{\compOp}{\bowtie}

\newcommand{\edge}{e}
\newcommand{\Edges}{E}
\newcommand{\longueflecheRel}[1]{\stackrel{#1}{\mapsto}}

\newcommand{\flecheRel}{{\rightarrow}}

\newcommand{\grandn}{{\mathbb N}}
\newcommand{\grandq}{{\mathbb Q}}
\newcommand{\grandqplus}{\grandq_{+}} %
\newcommand{\guard}{g}

\newcommand{\invariant}{I}
\newcommand{\K}{K}

\newcommand{\loc}{l} %
\newcommand{\locinit}{\loc\init}
\newcommand{\Loc}{L} %

\newcommand{\Param}{\mathbb{P}} %
\newcommand{\param}{p} %
\newcommand{\ParamCard}{M} %
\newcommand{\pval}{v} %

\newcommand{\R}{{\mathbb{R}}}
\newcommand{\grandrplus}{\R_{\geq 0}}

\newcommand{\sinit}{s\init} %
\newcommand{\somelocs}{T} %
\newcommand{\state}{\ensuremath{s}} %
\newcommand{\States}{S} %

\newcommand{\resets}{R}

\newcommand{\reset}[2]{\ensuremath{[#1]_{#2}}}
\newcommand{\valuate}[2]{\ensuremath{#2(#1)}}

\newcommand{\styleSched}[1]{\ensuremath{\mathsf{#1}}}
\newcommand{\EDF}{\ensuremath{\styleSched{EDF}}}
\newcommand{\FPS}{\ensuremath{\styleSched{FPS}}}
\newcommand{\RMS}{\ensuremath{\styleSched{RMS}}}
\newcommand{\SJF}{\ensuremath{\styleSched{SJF}}}
\newcommand{\TDMA}{\ensuremath{\styleSched{TDMA}}}

\newcommand{\stylealgo}[1]{\ensuremath{\textsf{#1}}}
\newcommand{\EFsynth}{\stylealgo{EFsynth}}

\usepackage{amsthm}
\theoremstyle{plain}
\theoremstyle{definition}
\newtheorem{definition}{Definition}
\newtheorem{example}{Example}

\theoremstyle{remark}
\usepackage{marginnote}
\ifdefined \VersionWithComments
	\newcommand{\marginX}{\marginnote{\huge{\quad\quad\textbf{!}\quad\quad}}}
	\newcommand{\ea}[1]{\mbox{}{\color{blue}\marginX{}\textbf{[\'Etienne}: #1]}}
	\newcommand{\instructions}[1]{{\color{red}\marginX{}\textbf{[Instructions: ``#1'']}}}
	\newcommand{\reviewer}[2]{\mbox{}{\color{red}\marginX{}\textbf{[Reviewer #1}: ``#2'']}}
	\newcommand{\todo}[1]{\mbox{}{\color{red}{\marginX{}\textbf{TODO}\ifx#1\\\else:\ \fi #1}}} %
\else
	\newcommand{\instructions}[1]{}
	\newcommand{\ea}[1]{}
	\newcommand{\reviewer}[2]{}
	\newcommand{\todo}[1]{}
\fi

\newcommand{\imitator}{\textsf{IMITATOR}}
\newcommand{\TimeForSys}{Time4sys}
\newcommand{\uppaal}{\textsc{Uppaal}}

\newcommand{\JawherTool}{Time4sys2IMITATOR}

\ifdefined \VersionWithComments
 	\definecolor{colorok}{RGB}{80,80,150}
\else
	\definecolor{colorok}{RGB}{0,0,0}
\fi

\newcommand{\eg}{\textcolor{colorok}{e.\,g.,}\xspace}
\newcommand{\ie}{\textcolor{colorok}{i.\,e.,}\xspace}
\newcommand{\st}{\textcolor{colorok}{s.t.}\xspace}

\begin{document}

\title{Formalizing Time4sys using parametric timed automata%
\thanks{%
	\LongVersion{%
		This is the author (and slightly extended) version of the manuscript of the same name published in the proceedings of the 13th International Symposium on Theoretical Aspects of Software Engineering (TASE 2019).
		The final version is available at \url{http://ieeexplore.ieee.org/}.
	}%
	This work is supported by the ASTREI project funded by the Paris Île-de-France Region,
	with the additional support of the ANR national research program PACS (ANR-14-CE28-0002)
	and
	ERATO HASUO Metamathematics for Systems Design Project (No.\ JPMJER1603), JST.
	}
}

\author{\IEEEauthorblockN{\'Etienne Andr\'e}
\IEEEauthorblockA{\textit{Universit\'e Paris 13, LIPN, CNRS, F-93430, Villetaneuse, France}\\
\textit{JFLI, CNRS}, %
Tokyo, Japan \\
\textit{National Institute of Informatics}, %
Tokyo, Japan
}
}

\LongVersion{
\pagestyle{plain}
}

\maketitle

\ifdefined \VersionWithComments
	\textcolor{red}{\textbf{This is the version with comments. To disable comments, comment out line~3 in the \LaTeX{} source.}}
\fi

\begin{abstract}
	Critical real-time systems must be verified to avoid the risk of dramatic consequences in case of failure.
	Thales developed an open formalism ``\TimeForSys{}'' to model real-time systems, with expressive features such as periodic or sporadic tasks, task dependencies, distributed systems, etc.
	However, \TimeForSys{} does not natively allow for a formal reasoning.
	In this work, we present a translation from \TimeForSys{} to (parametric) timed automata, so as to allow for a formal verification.
\end{abstract}


\section{Introduction}

Real-time systems combine concurrent behaviors with hard real-time constraints.
The correctness of real-time systems is crucial especially for \emph{critical} systems, the failure of which may cause irreparable damages.
A formal analysis of the system before its execution is therefore necessary.
Real-time systems are made of tasks (that can be periodic or not) to be executed on one or more processors.
Tasks are notably characterized by a relative deadline that stipulates how much time can be spent from the activation time of an instance of the task to the completion of that instance.
Often, a \emph{deadline miss} in a real-time system is an undesired behavior just as bad as a wrong result in a computation.
Each processor comes with a \emph{scheduling policy}, that decides what task should be executed.
Such policies can be preemptive, \ie{} may temporarily stop a task processing to move to a higher priority task.
Of utmost importance is the \emph{schedulability analysis}, that is: given a set of tasks to execute with their timing constraints (period, best case and worst case execution times, deadline) together with the scheduling policies, ensure that no deadline miss will occur.

Thales, a multinational company of 64,000 employees focusing on aerospace, defense and transportation, developed an internal tool for modeling real-time systems.
This tool was then recently made public under the name \TimeForSys{}, with its code made open source.\footnote{%
	\url{https://github.com/polarsys/time4sys}
}

\begin{figure*}[tb]
	\centering
	\includegraphics[width=.8\textwidth]{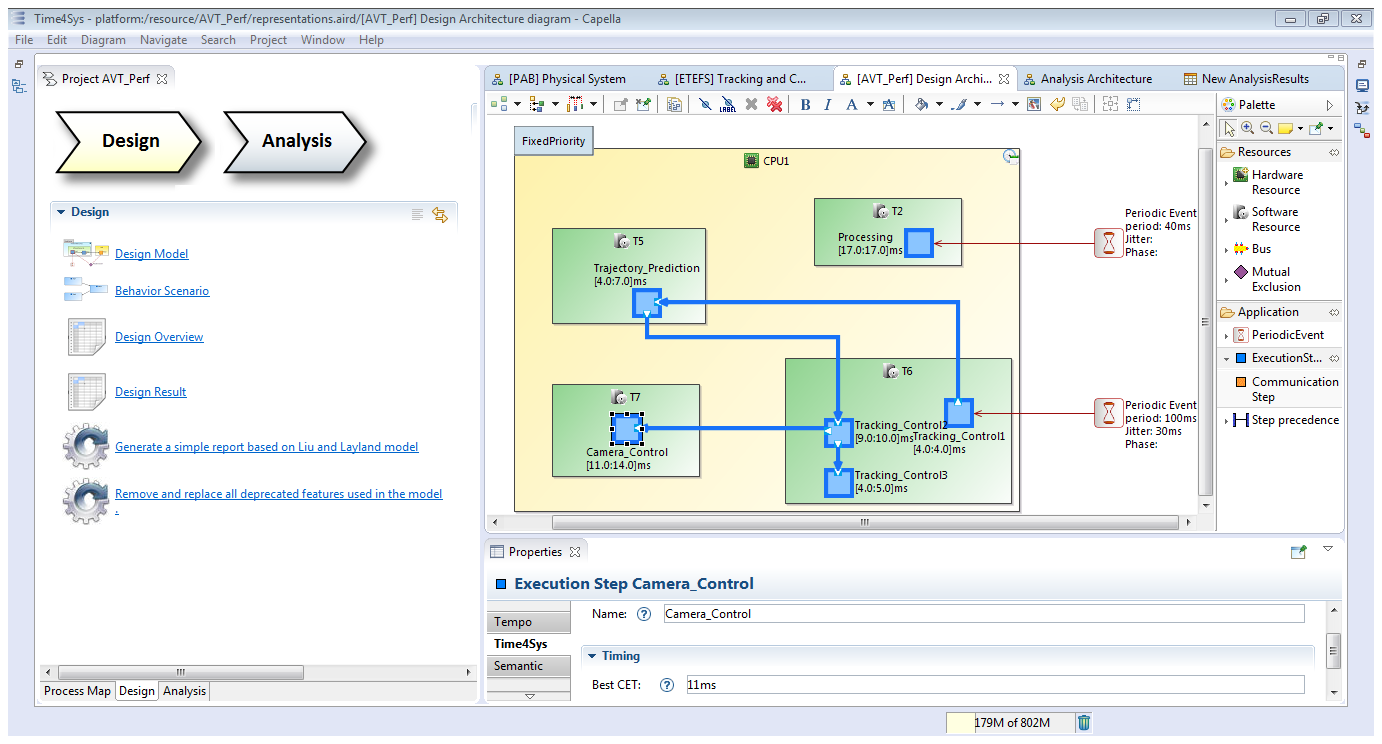}

	\caption{\TimeForSys{} graphical user interface}
	\label{figure:GUI}
\end{figure*}

\cref{figure:GUI} shows an example of \TimeForSys{}' graphical user interface, with one processor (``CPU1'', yellow square) on which 6~tasks (blue squares) are assigned.
Blue lines indicate task dependencies: for example, whenever task \texttt{Trajectory\_Prediction} is completed, an instance of \texttt{Tracking\_Control2} is activated.
Two tasks are periodically activated, \ie{} \texttt{Tracking\_Control1} (period: 100\,ms, jitter: 30\,ms), and \texttt{Processing} (period: 40\,ms, no jitter).

However, \TimeForSys{} mainly focuses on modeling, and not on verification.
While it is possible to design complex real-time systems using \TimeForSys{} and its graphical user interface (see \cref{figure:GUI}), no formal verification can be made.
As a consequence, transformation rules from a subset of the \TimeForSys{} syntax were proposed to the input format of tools such as Pegase\footnote{%
	\url{http://www.realtimeatwork.com/software/rtaw-pegase/}
}, MAST~\cite{GGPD01} or Cheddar~\cite{SLNM04}, which allow for some formal verification.
However, these tools only support some syntactic features: for example, Cheddar does not support task dependency constraints.
In 2015, the FMTV challenge\footnote{%
	``Formal Methods for Timing Verification Challenge'', in the WATERS workshop: \url{http://waters2015.inria.fr/}
}
made public a problem proposed by Thales, that consists in performing some verifications (and computations) on a real-time system model designed using \TimeForSys{}.
Due to \emph{uncertain} periods, \ie{} periodic tasks with a fixed but not completely known period, most solutions failed in computing the desired best-case and worst-case computation times.
Beside a solution that obtained the correct times using simulation, the only method able to compute these times in an exact fashion used parametric timed model checking~\cite{ALS15}.
Although modeled by \TimeForSys{}, the model of~\cite{ALS15} was formalized in an \emph{ad-hoc} manner to solve a given problem, while we propose here a systematic formalization.

\paragraph*{Contribution}
In this work, we propose a translation of the \TimeForSys{} syntax to an extension of timed automata~\cite{AD94}, a common formalism to verify systems involving concurrency and time.
In addition, we allow the use of parameters to cope for uncertainty by setting as target of our translation \emph{parametric} timed automata, that extend timed automata with unknown (or uncertain) constants~\cite{AHV93}.
In other words, some values such as periods or offsets can be completely unknown, or known with a limited certainty (\eg{} in a predefined interval).
The goal is to offer translation rules, so as to be able to analyze formally \TimeForSys{} models even in the presence of uncertainty.
Even further, the ultimate goal is to \emph{synthesize} the values of some timings constants seen as parameters that make the system schedulable.

\paragraph*{Related works}
\todo{R2: 1) The related work part should also discuss other techniques for schedulability analysis, apart from model checking-based approaches.}
Schedulability analysis of real-time systems using model checking was tackled in the past, with several works in the 1990s directly or indirectly targeting real-time systems (\eg{} \cite{WME92,AHV93,AD94,YMW97,CC99}).
In~\cite{AM02,AAM06}, schedulability analysis for some real-time systems is performed using timed automata~\cite{AD94} or stopwatch automata.\todo{more!}

When timing constants (periods, jitters, offsets, deadlines, etc.)\ become uncertain or unknown, schedulability analysis becomes much harder.
A method is to use parametric timed formalisms such as parametric timed automata~\cite{AHV93}.
In~\cite{CPR08} parametric automata are used to \emph{synthesize} values for offsets, periods, deadlines or computation times for which the system is schedulable.
While the problem is in general undecidable, a decidable subclass is exhibited.
In~\cite{FLMS12}, the authors use stopwatch automata to perform robust schedulability analysis, \ie{} where some of the timing constants can vary while preserving the schedulability.
In~\cite{SSLAF13}, we proposed a method to synthesize values of values such as deadlines and periods for which schedulability is ensured for a real-time system modeled using a set of pipelines: while the method is slower than an analytical method and MAST~\cite{GGPD01}, our method is also more complete.
Despite some \emph{ad-hoc} experiments, no automatic transformation from an existing formalism to parametric timed automata was proposed.

For uniprocessor real-time systems only, (parametric) \emph{task automata} offer a more compact representation than (parametric) timed automata~\cite{FKPY07,NWY99,Andre17FMICS}.

Finally, Time4sys was partially translated to parametric time Petri nets~\cite{LRST09}.

\paragraph*{Outline}
We present \TimeForSys{} in \cref{section:time4sys}.
We recall the syntax and semantics of parametric timed automata in \cref{section:preliminaries}.
We present our main transformation in \cref{section:transformation},
and we report experiments as a proof of concept in \cref{section:experiments}.
Finally, we conclude and outline perspectives in \cref{section:conclusion}.

\section{\TimeForSys{}}\label{section:time4sys}

\TimeForSys{}\footnote{%
	\url{https://fed4sae.eu/industrial-platforms/time4sys-thales/}
}
is a graphical representation of real-time systems designed and released by Thales.

\TimeForSys{} has a formally established syntax given in the form of a metamodel.
\TimeForSys{} uses a subset of the MARTE OMG standard~\cite{MARTE} as a basis to represent a synthetic view of the system design model that captures all elements, data and properties impacting the system timing behavior and required to perform timing verification (\eg{} tasks mapping on processors, communication links, execution times, scheduling parameters, etc.).

By using MDE (model-driven engineering) settings, \TimeForSys{} is being developed as an Eclipse Polarsys plugin, and comes with a user-friendly graphical interface.

While the semantics of real-time systems modeled in \TimeForSys{} is clear (perhaps up to some minor aspects),
	\TimeForSys{} does not natively allow itself for any verification, and relies on external tools such as MAST~\cite{GGPD01} or Cheddar~\cite{SLNM04}.

\begin{figure*}
	\centering
	\begin{tikzpicture}[scale=.7, xscale=1, yscale=.7]
		\footnotesize

		\draw[thick] (-1, 0) --++ (0, 2.75);
		\draw[thick,->] (-1, 2) -- (17.5,2) node[right] {CPU1};
		\draw[thick,->] (-1, 1) -- (17.5,1) node[right] {CPU2};
		\draw[thick,->] (-1, 0) -- (17.5,0) node[right] {CPU1};

		\foreach \x in {0, 1, ..., 17} %
			\draw[very thin] (\x, 2.5) -- (\x, -.1) node [below] {\scriptsize $\x$};

		\draw[draw=none,fill=blue!50] (0, 0) rectangle (5, .5);
		\node at (2.5, .25) {T5};
		
		\draw[draw=none,fill=red!50] (5, 0) rectangle (9.4, .5);
		\node at (5/2 + 9.4/2, .25) {T1};
		
		\draw[draw=none,fill=blue!50] (9.4, 0) rectangle (11.4, .5);
		\node at (10.4, .25) {T5};
		
		\draw[draw=none,fill=red!50] (15, 0) rectangle (17, .5);
		\node at (16, .25) {T1};

		\draw[draw=none,fill=orange!50] (9.4, 1) rectangle (10.4, 1.5);
		\node at (9.7, 1.25) {T2};
		
		\draw[draw=none,fill=purple!50] (10.4, 1) rectangle (12.2, 1.5);
		\node at (10.9, 1.25) {T7};
		
		\draw[draw=none,fill=gray!50] (12.2, 1) rectangle (13.2, 1.5);
		\node at (12.7, 1.25) {T4};
		
		\draw[draw=none,fill=purple!50] (13.2, 1) rectangle (17, 1.5);
		\node at (13.2/2 + 17/2, 1.25) {T7};

		\draw[draw=none,fill=green!50] (4, 2) rectangle (10, 2.5);
		\node at (6.5, 2.25) {T6};
		
		\draw[draw=none,fill=yellow!50] (10.4, 2) rectangle (12.2, 2.5);
		\node at (10.4/2 + 12.2/2, 2.25) {T3};

		\draw[activation, draw=blue] (0, 0) --++ (0,.75);
		\draw[activation, draw=red] (5, 0) --++ (0,.75);
		\draw[activation, draw=red] (15, 0) --++ (0,.75);
		
		\draw[activation, draw=orange] (9.4, 1) --++ (0,.75);
		\draw[activation, draw=purple] (10, 1) --++ (0,.75);
		\draw[activation, draw=gray] (12.2, 1) --++ (0,.75);
		
		\draw[activation, draw=green] (4, 2) --++ (0,.75);
		\draw[activation, draw=yellow] (10.4, 2) --++ (0,.75);

	\end{tikzpicture}

	\caption{The beginning of a possible execution of the system in \cref{figure:example:Time4sys}}
	\label{figure:chronogramme}
\end{figure*}
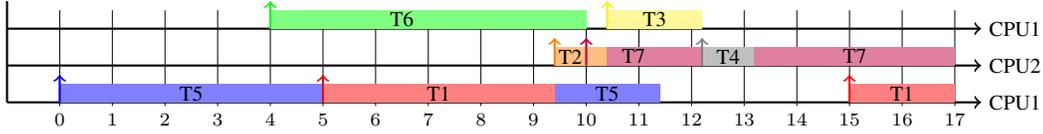

\paragraph*{Supported features}
\TimeForSys{} allows to model processors (``hardware resources'') and tasks (``software resources''), with their execution times (possibly in the form an interval) and their deadline.
In addition, tasks can be activated periodically or sporadically, possibly with a minimum inter-arrival time.
Periodic tasks can be subject to a jitter~$j$: that is, the $i$th instance of a task with period~$p$ will occur in $[(i-1) \times p - j , (i-1) \times p + j]$.
In addition, tasks can be subject to an offset~$o$: in that case, the $i$th instance of a task with period~$p$ will occur at time $(i-1) \times p + o$.\todo{macros, variable $p$, etc.}

Tasks usually come with a relative \emph{deadline}: that is, every time a new task instance is activated, this instance must complete its execution within this deadline.
Otherwise, a \emph{deadline miss} occurs, and the system is not schedulable.
The schedulability analysis consists in formally checking that, for any execution consistent with the task model (periods, offsets, deadlines, execution times, scheduling policy…), no deadline miss will ever occur.

\todo{ajout schéma du genre de \cite[fig.2]{FBGLP10}}

\TimeForSys{} supports both uniprocessor and multi-processor real-time systems.
Task dependencies are supported, including between different processors; that is, the completion of a task on a given processor can trigger the activation of a task on the same or on another processor.
This is perhaps one of the key aspects of \TimeForSys{}, as these dependencies are not always supported by existing schedulability analysis techniques.

Various scheduling policies are supported by \TimeForSys{}, including:
\begin{itemize}
	\item fixed priority (\FPS{}): each task on a given processor comes with a given static priority and, upon completion of a task instance, the highest priority task instance in the waiting queue is selected for execution;
	\item preemptive \FPS{}: a version of \FPS{} where a lower priority task can be temporarily stopped to execute a newly activated instance of a higher priority task;
	\item (preemptive) \RMS{}: \FPS{} where tasks with shorter periods necessarily have a higher priority;
	\item shortest job first (\SJF{}): the shortest task instance is selected first;
	\item Time-division multiple access (\TDMA{}) and Round Robin: tasks are allocated slots in a cyclic manner to execute (part of) their instances, one after the other.
\end{itemize}

\begin{figure*}
	\centering
	\includegraphics[width=.8\textwidth]{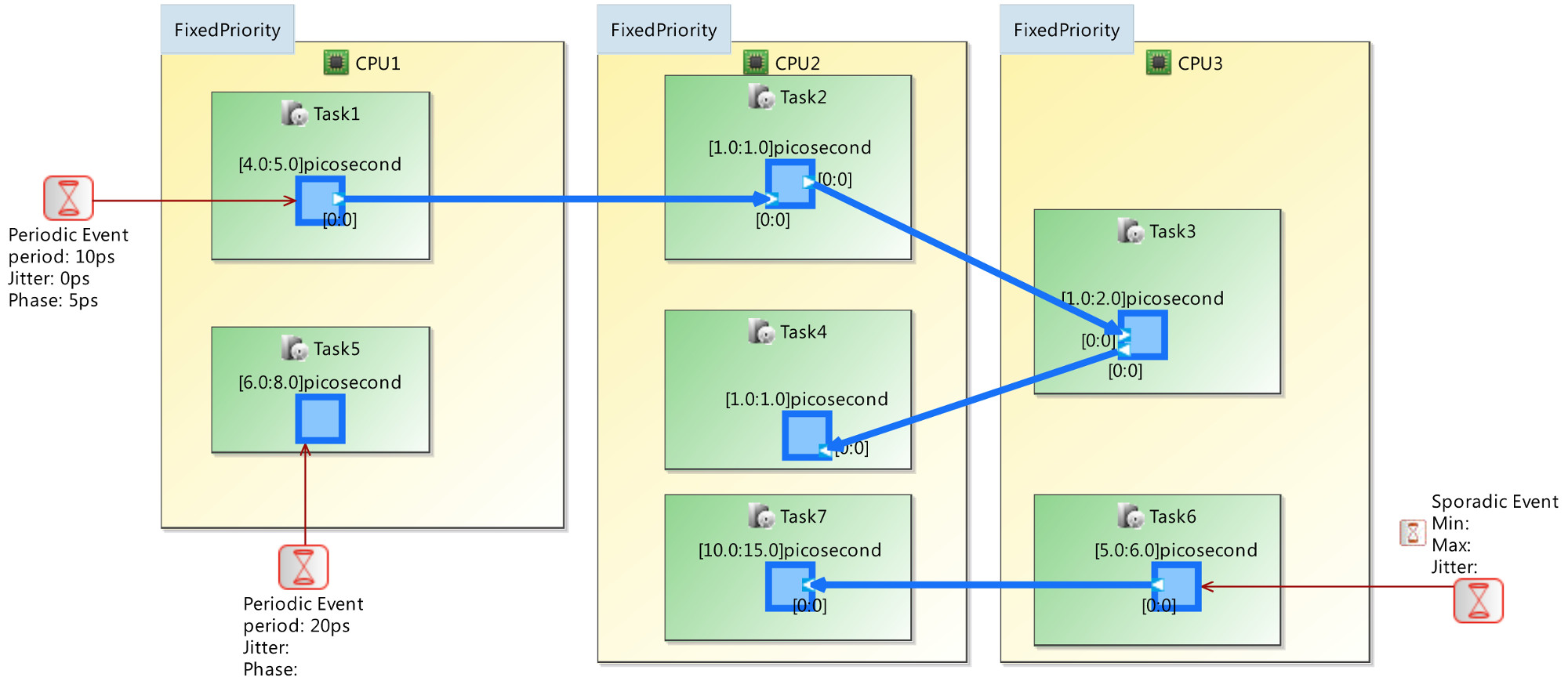}

	\caption{Example of \TimeForSys{} model}
	\label{figure:example:Time4sys}
\end{figure*}

\begin{example}
	Consider the real-time system in \cref{figure:example:Time4sys} modeled using \TimeForSys{} (the $[0;0]$ intervals should be disregarded).
	This system features seven tasks to be executed on three processors.
	Processors are depicted in yellow boxes, while tasks are depicted using blue squares (green boxes can be disregarded in our framework).
	Blue lines represent task dependencies, while red lines denote sporadic or periodic activations.
	Tasks $T_1$ and $T_5$ are periodic, $T_6$ is sporadic, while others are activated by task dependencies.
	Only task~$T_1$ is subject to an offset (called ``phase'' in \TimeForSys{}).
	All three processors use the \FPS{} scheduling policy.
	Assume that periodic tasks (Task~1 and Task~6) have deadlines equal to their period (\ie{} 10 and 20, respectively).

	Let us consider the beginning of a possible execution of this system (given in \cref{figure:chronogramme}).
	At system start ($t = 0$), Task~5 will be activated; as Task~1 is not activated (due to the offset of~5), Task~5 (although of lower priority) executes on CPU1.
	On CPU3, Task~6 may be activated anytime, as this is a sporadic task.
	Let us assume Task~6 is not activated at $t = 0$.
	At $t = 0$, no other task is activated and therefore only CPU1 is active, and both CPU2 and CPU3 are idle.
	At $t = 4$, Task~6 is activated, and starts on CPU3.
	Then, at $t = 5$ (which is the offset of Task~1), an instance of Task~1 is activated; as Task~1 has higher priority than Task~5, Task~5 is \emph{preempted}, and Task~1 starts executing.
	At $t=9.4$, Task~1 finishes its computation (recall that its computation time lies in $[4,5]$).
	Then, at that time, an instance of Task~2 is activated by the completion of Task~1; as CPU2 is idle, it starts to execute immediately.
	In parallel, the computation of Task~5 resumes on CPU1; assuming this execution of Task~5 lasts 7 time units (in interval $[6,8]$), then this instance will be completed at $t = 11.4$.
	At $t=10$, Task~6 completes, triggering the activation of Task~7 on CPU2---which must however wait the end of T2 since it is of lower priority.
	And so on.\ea{finish commenting?}
	
	Now observe that, if the deadline of Task~5 was 11 (instead of~20), then at $t = 11$, a deadline miss would occur for Task~5, as the execution of its instance is not completed by its deadline\LongVersion{ (see \cref{appendix:miss1} for details)}.
	Or, alternatively, if the deadline of Task~5 was 20 but the worst case execution time of Task~1 was 15, then at $t=20$, Task~5 has still not completed, and again a deadline miss occurs\LongVersion{ (see \cref{appendix:miss2} for details)}.
	
	This justifies the study of two problems for \TimeForSys{} models:
	\begin{itemize}
		\item being able to verify that a system is schedulable;
		\item being able to exhibit suitable values (notably for execution times and deadlines) for which the system is schedulable.
	\end{itemize}
\end{example}
\section{Parametric timed automata}\label{section:preliminaries}

Let $\grandn$, $\grandqplus$ and $\grandrplus$ denote the set of non-negative integers, non-negative rationals and non-negative reals, respectively.

We assume a set~$\Clock = \{ \clock_1, \dots, \clock_\ClockCard \} $ of \emph{clocks}, \ie{} real-valued variables that evolve at the same rate.
A clock valuation is a function
$\clockval : \Clock \rightarrow \grandrplus$.
We write $\ClocksZero$ for the clock valuation assigning $0$ to all clocks.
Given $d \in \grandrplus$, $\clockval + d$ \ShortVersion{is}\LongVersion{denotes the valuation} \st{} $(\clockval + d)(\clock) = \clockval(\clock) + d$, for all $\clock \in \Clock$.
Given $\resets \subseteq \Clock$, we define the \emph{reset} of a valuation~$\clockval$, denoted by $\reset{\clockval}{\resets}$, as follows: $\reset{\clockval}{\resets}(\clock) = 0$ if $\clock \in \resets$, and $\reset{\clockval}{\resets}(\clock)=\clockval(\clock)$ otherwise.

We assume a set~$\Param = \{ \param_1, \dots, \param_\ParamCard \} $ of \emph{parameters}\LongVersion{, \ie{} unknown constants}.
A parameter {\em valuation} $\pval$ is\LongVersion{ a function}
$\pval : \Param \rightarrow \grandqplus$.
We assume ${\compOp} \in \{<, \leq, =, \geq, >\}$.
A \emph{guard}~$\guard$ is a constraint over $\Clock \cup \Param$ defined by a conjunction of inequalities of the form $\clock \compOp d$, 
or $\clock \compOp \param$ with $\clock \in \Clock$, $d \in \grandn$ and $\param \in \Param$.
Given~$\guard$, we write~$\clockval\models\pval(\guard)$ if %
the expression obtained by replacing each~$\clock$ with~$\clockval(\clock)$ and each~$\param$ with~$\pval(\param)$ in~$\guard$ evaluates to true.

Parametric timed automata (PTA) extend timed automata~\cite{AD94} with parameters within guards and invariants in place of integer constants~\cite{AHV93}.

\begin{definition}[PTA]\label{def:uPTA}
	A PTA $\A$ is a tuple \mbox{$\A = (\Actions, \Loc, \locinit, \Clock, \Param, \invariant, \Edges)$}, where:
	\begin{ienumeration}
		\item $\Actions$ is a finite set of synchronization actions,
		\item $\Loc$ is a finite set of locations,
		\item $\locinit \in \Loc$ is the initial location,
		\item $\Clock$ is a finite set of clocks,
		\item $\Param$ is a finite set of parameters,
		\item $\invariant$ is the invariant, assigning to every $\loc\in \Loc$ a guard $\invariant(\loc)$,
		\item $\Edges$ is a finite set of edges  $\edge = (\loc,\guard,\action,\resets,\loc')$
		where~$\loc,\loc'\in \Loc$ are the source and target locations, $\action \in \Actions$, $\resets\subseteq \Clock$ is a set of clocks to be reset, and $\guard$ is a guard.
	\end{ienumeration}
\end{definition}

Given\LongVersion{ a parameter valuation}~$\pval$, we denote by $\valuate{\A}{\pval}$ the non-parametric structure where all occurrences of a parameter~$\param_i$ have been replaced by~$\pval(\param_i)$.
\LongVersion{%
	We denote as a \emph{timed automaton} any structure $\valuate{\A}{\pval}$, by assuming a rescaling of the constants: by multiplying all constants in $\valuate{\A}{\pval}$  by their least common denominator, we obtain an equivalent (integer-valued) TA\LongVersion{, as defined in \cite{AD94}}.
}

\begin{example}
	An example of PTA is given in \cref{figure:translation:periodic} (page~\pageref{figure:translation:periodic}).
	It features two locations $\styleloc{l1}$ and~$\styleloc{l2}$, one synchronization action $\styleact{actT}$, one clock~$\styleclock{xactT}$, and two parameters $\styleparam{TPeriod}$ and $\styleparam{TOffset}$.
\end{example}

\LongVersion{%
	Let us now recall the concrete semantics of TA.
}

\begin{definition}[Semantics of a TA]
	Given a PTA $\A = (\Actions, \Loc, \locinit, \Clock, \Param, \invariant, \Edges)$,
	and a parameter valuation~\(\pval\),
	the semantics of $\valuate{\A}{\pval}$ is given by the timed transition system (TTS) $(\States, \sinit, \flecheRel)$, with
	\begin{itemize}
		\item $\States = \{ (\loc, \clockval) \in \Loc \times \grandrplus^\ClockCard \mid \clockval \models \valuate{\invariant(\loc)}{\pval} \}$, %
		\LongVersion{\item }$\sinit = (\locinit, \ClocksZero) $,
		\item  $\flecheRel$ consists of the discrete and (continuous) delay transition relations:
		\begin{ienumeration}
			\item discrete transitions: $(\loc,\clockval) \longueflecheRel{\edge} (\loc',\clockval')$, %
				if $(\loc, \clockval) , (\loc',\clockval') \in \States$, and there exists $\edge = (\loc,\guard,\action,\resets,\loc') \in \Edges$, such that $\clockval'= \reset{\clockval}{\resets}$, and $\clockval\models\pval(\guard$).
			\item delay transitions: $(\loc,\clockval) \longueflecheRel{d} (\loc, \clockval+d)$, with $d \in \grandrplus$, if $\forall d' \in [0, d], (\loc, \clockval+d') \in \States$.
		\end{ienumeration}
	\end{itemize}
\end{definition}

Given a TA~$\valuate{\A}{\pval}$ with concrete semantics $(\States, \sinit, \flecheRel)$, we refer to the states of~$\States$ as the \emph{concrete states} of~$\valuate{\A}{\pval}$.
A \emph{run} of~$\valuate{\A}{\pval}$ is an alternating sequence of concrete states of $\valuate{\A}{\pval}$ and pairs of edges and delays starting from the initial state $\sinit$ of the form
$\state_0, (\edge_0, d_0), \state_1, \cdots$
with
$i = 0, 1, \dots$, $\edge_i \in \Edges$, $d_i \in \grandrplus$ and $(\state_i , \edge_i , \state_{i+1}) \in \flecheRel$.
Given\LongVersion{ a state}~$\state=(\loc, \clockval)$, we say that $\state$ is reachable in~$\valuate{\A}{\pval}$ if $\state$ appears in a run of $\valuate{\A}{\pval}$.
By extension, we say that $\loc$ is reachable; and by extension again, given a set~$\somelocs$ of locations, we say that $\somelocs$ is reachable if there exists $\loc \in \somelocs$ such that $\loc$ is reachable in~$\valuate{\A}{\pval}$.

\subsection{Reachability synthesis}

We will use here reachability synthesis to perform parametric schedulability analysis.
That is, the system will be schedulable for valuations of the PTA for which the valuated TA does not reach a set of given locations modeling deadline misses.

The procedure synthesizing valuations for which a set of locations is (un)reachable is called \EFsynth{}: it takes as input a PTA~$\A$ and a set of target locations~$\somelocs$, and attempts to synthesize all parameter valuations~$\pval$ for which~$\somelocs$ is reachable in~$\valuate{\A}{\pval}$.
\EFsynth{} was given and formalized in \eg{} \cite{JLR15} and may not terminate, but if it terminates, then its result is exact (sound and complete).
\LongVersion{%
	\EFsynth{} traverses the \emph{parametric zone graph} of~$\A$, which is a potentially infinite extension of the well-known zone graph of TAs (see, \eg{} \cite{AS13,JLR15}\LongVersion{ for a formal definition}).
}

\subsection{\imitator{}}

\imitator{}~\cite{AFKS12} is a parametric timed model checker supporting networks of parametric timed automata extended with various features such as stopwatches (\ie{} the ability to stop the elapsing of some clocks, allowing to model preemption in real-time systems), synchronization using synchronization actions, global discrete rational-valued variables that can read and modified in guards, etc.
In the following, we refer to these extended PTAs as PTAs.
Our translation takes advantage of the features offered by \imitator{}, notably stopwatches and action synchronization.
\LongVersion{

}%
\imitator{} was used in the past to verify several industrial case studies (\eg{} \cite{FLMS12,ALS15}).

\section{Translating \TimeForSys{} into PTAs}\label{section:transformation}

In this section, we present our translation from the syntax of \TimeForSys{} into parametric timed automata (extended with stopwatches).
We use the real-time system in \cref{figure:example:Time4sys} to exemplify our translation.

\subsection{Assumptions}\label{ss:assumptions}

We make the following assumptions on the \TimeForSys{} models we aim at analyzing:
\begin{enumerate}
	\item We allow only one-to-one task dependencies; that is, a given task upon completion can activate only a single task.
		This is not the case of \cref{figure:GUI} as \texttt{Tracking\_control2} activates two tasks; in contrast, this assumption is satisfied in \cref{figure:example:Time4sys}.
	\item For periodic tasks, the  deadline is equal to the period.
	\item No jitters are allowed.
\end{enumerate}

How to lift these assumptions is discussed in \cref{section:conclusion}.

\subsection{General translation scheme}

We will build a network of PTAs synchronized on synchronization actions.
By default, we assume the model is fully parameterized, \ie{} all timing constants (deadlines, periods, offsets, jitters…)\ are parameters in the sense of PTAs' unknown constants.
Assigning these parameters to a fixed value can be done trivially in \imitator{}, and makes the subsequent analysis simpler.
For example, in \cref{figure:example:Time4sys}, we should set $\styleparam{T1period} := 10\,ps$.

In addition, we use the following same convention as in the literature, as well as in \imitator{}: different objects (clocks, parameters, variables, actions---but not locations) with the same name in different PTAs refer to \emph{shared} objects.
For example, a clock named $\clock$ used in two different PTAs is the \emph{same} clock.
However, two locations $\loc_1$ in two different PTAs are (naturally) two different locations.

The various PTAs will be synchronized using synchronization actions.
Given a task~$T$, two actions will be used: \styleact{actT}, denoting the activation of an instance of~$T$, and \styleact{finT} to denote its completion.
Note that, thanks to our assumption of deadlines equal to periods, not more than one instance of a task is present (executing or waiting) at a given time: indeed, if a second instance of a task is activated, this would mean that the first one did not complete within its period.
\subsection{Task activation patterns}

We consider two main kinds of activation patterns: periodic tasks and sporadic tasks.

\paragraph{Periodic tasks}

\begin{figure*}[tb]
	\centering
	\footnotesize
	\begin{subfigure}[b]{.49\textwidth}
		
		\centering
			\begin{tikzpicture}[scale=3, auto, ->, >=stealth']
 
		\node[location, initial, fill=loccolor1] at (0,0) (l1) {\styleloc{l1}};
		\node [invariant,below] at (l1.south) {\begin{tabular}{@{} c @{\ } c@{} }& $ \styleclock{xactT} \leq \styleparam{TOffset}$\\\end{tabular}};
 
		\node[location, fill=loccolor2] at (1, 0) (l2) {\styleloc{l2}};
		\node [invariant,below] at (l2.south) {\begin{tabular}{@{} c @{\ } c@{} }& $ \styleclock{xactT} \leq \styleparam{TPeriod}$\\\end{tabular}};

		\path (l1) edge node[above]{\begin{tabular}{@{} c @{\ } c@{} }
		& $ \styleclock{xactT} = \styleparam{TOffset}$\\
		 & $\styleact{actT}$\\
		 & $\styleclock{xactT}:=0$\\
		\end{tabular}} (l2);

		\path (l2) edge[loop above] node[below right]{\begin{tabular}{@{} c @{\ } c@{} }
		& $ \styleclock{xactT} = \styleparam{TPeriod}$\\
		 & $\styleact{actT}$\\
		 & $\styleclock{xactT}:=0$\\
		\end{tabular}} (l2);
	\end{tikzpicture}
		
		\caption{Periodic task}
		\label{figure:translation:periodic}
	\end{subfigure}
	\hfill{}
	\begin{subfigure}[b]{.49\textwidth}
	
		\centering
		
			\begin{tikzpicture}[scale=3, auto, ->, >=stealth']
 
		\node[location, initial, fill=loccolor1] at (0,0) (l1) {\styleloc{l1}};
		\node [invariant,below] at (l1.south) {\begin{tabular}{@{} c @{\ } c@{} }& $ \styleclock{xactT} \leq \styleparam{TOffset}$\\\end{tabular}};
 
		\node[location, fill=loccolor2] at (1,0) (l2) {\styleloc{l2}};

		\path (l1) edge node[above]{\begin{tabular}{@{} c @{\ } c@{} }
		& $ \styleclock{xactT} = \styleparam{TOffset}$\\
		 & $\styleact{actT}$\\
		 & $\styleclock{xactT}:=0$\\
		\end{tabular}} (l2);

		\path (l2) edge[loop above] node[below right]{\begin{tabular}{@{} c @{\ } c@{} }
		& $ \styleclock{xactT} \geq \styleparam{TIAT}$\\
		 & $\styleact{actT}$\\
		 & $\styleclock{xactT}:=0$\\
		\end{tabular}} (l2);
	\end{tikzpicture}
		
		\caption{Sporadic task}
		\label{figure:translation:sporadic}
	\end{subfigure}
	
	\caption{Translating activation patterns}
	\label{figure:example:TODO}
\end{figure*}
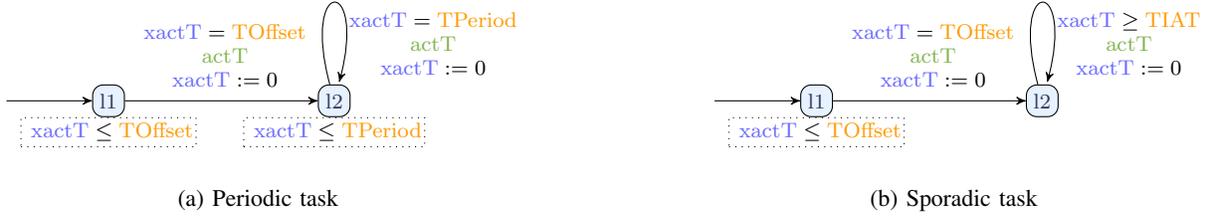

Given a task~$T$ with period \styleparam{TPeriod} and offset \styleparam{TOffset}, we create a PTA with a local clock~$\styleclock{xactT}$ synchronizing on action~\styleact{actT}, given in \cref{figure:translation:periodic}.
The first location is used to encode the offset at the beginning, while the second one is used to encode the periodic behavior.
Initially, in \styleloc{l1}, the system must wait \styleparam{TOffset} time units before starting the periodic behavior, according to the definition of the offset.
Then, after exactly \styleparam{TOffset} time units (encoded by the guard $\styleclock{xactT} = \styleparam{TOffset}$), the PTA activates task~$T$ using the synchronization action \styleact{actT} and moves to the second location.
There, every exactly \styleparam{TPeriod} time units (encoded by the guard $\styleclock{xactT} = \styleparam{TPeriod}$), the PTA activates task~$T$.

\paragraph{Sporadic tasks}

Sporadic tasks are characterized by their minimum inter-arrival time.
The encoding of sporadic task~$T$ with minimum inter-arrival time~\styleparam{TIAT} is very similar to the case of a periodic task, and is given in \cref{figure:translation:sporadic}.
The only differences are the absence of invariant in the lower location, and the fact that the activation is not anymore when $ \styleclock{xactT} = \styleparam{TPeriod}$, but when $\styleclock{xactT} \geq \styleparam{TIAT}$.

\subsection{Task precedences}\label{ss:taskchains}

We now encode task precedences.
Recall that, although this syntactic feature is intuitive (``upon completion, a given task triggers the activation of another task''), many tools do not support them; for example, Cheddar does not support task such dependency constraints.
Due to our assumption banning multiple precedences, such task precedences can be seen as \emph{task chains}, that are reminiscent of the pipelines that we modeled in~\cite{SSLAF13} using PTAs.
For example, in \cref{figure:example:Time4sys}, there are three task chains:
\begin{ienumeration}
	\item $T_1 \rightarrow T_2 \rightarrow T_3 \rightarrow T_4$;
	\item $T_6 \rightarrow T_7$; and
	\item $T_5$.
\end{ienumeration}

Translating these task chains is done by constraining the order between the various task activations and completions.
However, since the total execution time of the whole task chain can be larger than the first task's period, a single automaton is not sufficient.
Therefore, we ``split'' the chain, and encode each dependency as a PTA; as usual, these PTAs will be synchronized on the synchronization actions.

\begin{figure}[tb]
	\centering
	\footnotesize
	\begin{subfigure}[b]{\columnwidth}
		\centering
		
			\begin{tikzpicture}[scale=2, auto, ->, >=stealth']
 
		\node[location, initial, fill=loccolor1] at (0,0) (l1) {\styleloc{l1}};
 
		\node[location, fill=loccolor2] at (1, 0) (l2) {\styleloc{l2}};
 
		\node[location, urgent, fill=loccolor3] at (2, 0) (l3) {\styleloc{U: l3}};
 
		\node[location, fill=loccolor4] at (3, 0) (l4) {\styleloc{l1}};

		\path (l1) edge node{\begin{tabular}{@{} c @{\ } c@{} }
		 & $\styleact{actT1}$\\
		\end{tabular}} (l2);

		\path (l2) edge node{\begin{tabular}{@{} c @{\ } c@{} }
		 & $\styleact{finT1}$\\
		\end{tabular}} (l3);

		\path (l3) edge node{\begin{tabular}{@{} c @{\ } c@{} }
		 & $\styleact{actT2}$\\
		\end{tabular}} (l4);
	\end{tikzpicture}

	
		
		\caption{Tasks $T_1$ and $T_2$}
		\label{figure:translation:taskchains12}
	\end{subfigure}
	
	\begin{subfigure}[b]{\columnwidth}
		\centering
		
			\begin{tikzpicture}[scale=2, auto, ->, >=stealth']
 
		\node[location, initial, fill=loccolor1] at (0,0) (l1) {\styleloc{l1}};
 
		\node[location, fill=loccolor2] at (1, 0) (l2) {\styleloc{l2}};
 
		\node[location, urgent, fill=loccolor3] at (2, 0) (l3) {\styleloc{U: l3}};
 
		\node[location, fill=loccolor4] at (3, 0) (l4) {\styleloc{l1}};

		\path (l1) edge node{\begin{tabular}{@{} c @{\ } c@{} }
		 & $\styleact{actT2}$\\
		\end{tabular}} (l2);

		\path (l2) edge node{\begin{tabular}{@{} c @{\ } c@{} }
		 & $\styleact{finT2}$\\
		\end{tabular}} (l3);

		\path (l3) edge node{\begin{tabular}{@{} c @{\ } c@{} }
		 & $\styleact{actT3}$\\
		\end{tabular}} (l4);
	\end{tikzpicture}

	
		
		\caption{Tasks $T_2$ and $T_3$}
		\label{figure:translation:taskchains23}
	\end{subfigure}
	
	\begin{subfigure}[b]{\columnwidth}
		\centering
		
			\begin{tikzpicture}[scale=2, auto, ->, >=stealth']
 
		\node[location, initial, fill=loccolor1] at (0,0) (l1) {\styleloc{l1}};
 
		\node[location, fill=loccolor2] at (1, 0) (l2) {\styleloc{l2}};
 
		\node[location, urgent, fill=loccolor3] at (2, 0) (l3) {\styleloc{U: l3}};
 
		\node[location, fill=loccolor4] at (3, 0) (l4) {\styleloc{l1}};

		\path (l1) edge node{\begin{tabular}{@{} c @{\ } c@{} }
		 & $\styleact{actT3}$\\
		\end{tabular}} (l2);

		\path (l2) edge node{\begin{tabular}{@{} c @{\ } c@{} }
		 & $\styleact{finT3}$\\
		\end{tabular}} (l3);

		\path (l3) edge node{\begin{tabular}{@{} c @{\ } c@{} }
		 & $\styleact{actT4}$\\
		\end{tabular}} (l4);
	\end{tikzpicture}

	
		
		\caption{Tasks $T_3$ and $T_4$}
		\label{figure:translation:taskchains34}
	\end{subfigure}
	
	\caption{Translating task chain $T_1 \rightarrow T_2 \rightarrow T_3 \rightarrow T_4$}
	\label{figure:translation:taskchains}
\end{figure}
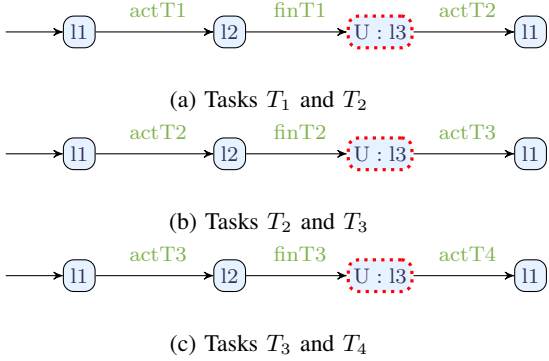

We give the translation of the task chain ``$T_1 \rightarrow T_2 \rightarrow T_3 \rightarrow T_4$'' in the three PTAs given in \cref{figure:translation:taskchains}.
The three locations named \styleloc{l3} are urgent locations (preceded by ``\styleloc{U:}''), \ie{} locations where time cannot elapse: that is, in \cref{figure:translation:taskchains12}, the activation of task $T_2$ comes immediately after the completion of~$T_1$.

\subsection{Scheduling policies}\label{ss:scheduling}

The scheduler of each CPU is translated into a PTA according to its scheduling policy.
We focus on preemptive \FPS{} as this is one of the most widely used policies, and arguably one of the most complex to encode.
Recall that \FPS{} always executes the highest priority task (all tasks of a given processor have a different priority); preemption allows to stop the ongoing computation to move to a higher priority incoming task.
\RMS{} is a particular case of \FPS{} where a shorter period is necessarily assigned a higher priority (not necessarily the case in \FPS{}).

\paragraph*{Encoding \FPS{}}

\begin{figure*}[tb]
	\centering
	\scriptsize

				\scalebox{.95}{
	\begin{tikzpicture}[scale=3, auto, ->, >=stealth']
 
		\node[location, initial, fill=loccolor1] at (0,.8) (idle) {\styleloc{idle}};
		\node [invariant,above left,align=center] at (idle.west) {stop(\styleclock{xexecT1},\\ \styleclock{xexecT5}};
 
		\node[location, fill=loccolor2] at (1.5,1.6) (execT1) {\styleloc{execT1}};
		\node [invariant,above] at (execT1.north) {\begin{tabular}{@{} c @{\ } c@{} }& $ \styleparam{T1WCET} \geq \styleclock{xexecT1}$\\ & stop(\styleclock{xexecT5})\end{tabular}};
 
		\node[location, fill=loccolor3] at (1.5,.8) (execT1waitT5) {\styleloc{execT1waitT5}};
		\node [invariant,below right] at (execT1waitT5.east) {\begin{tabular}{@{} c @{\ } c@{} }& $ \styleparam{T1WCET} \geq \styleclock{xexecT1}$\\ & stop(\styleclock{xexecT5})\end{tabular}};
 
		\node[location, fill=loccolor4] at (1.5,0) (execT5) {\styleloc{execT5}};
		\node [invariant,below] at (execT5.south) {\begin{tabular}{@{} c @{\ } c@{} }& $ \styleparam{T5WCET} \geq \styleclock{xexecT5}$\\ & stop(\styleclock{xexecT1})\end{tabular}};
 
		\node[location, fill=loccolor5] at (3.5,.8) (DeadlineMissed) {\styleloc{DeadlineMissed}};

		\path (idle) edge[bend angle=20,bend right] node{\begin{tabular}{@{} c @{\ } c@{} }
		 & $\styleact{actT1}$\\
		 & $\styleclock{xexecT1}:=0$\\
		\end{tabular}} (execT1);

		\path (idle) edge[bend angle=20,bend left] node[below left]{\begin{tabular}{@{} c @{\ } c@{} }
		 & $\styleact{actT5}$\\
		 & $\styleclock{xexecT5}:=0$\\
		\end{tabular}} (execT5);

		\path (execT1) edge[bend right] node[above left]{\begin{tabular}{@{} c @{\ } c@{} }
		& $ \styleclock{xexecT1} \geq \styleparam{T1BCET}$\\
		 & $\styleact{finT1}$\\
		 & $\styleclock{xactT1}:=0$\\
		\end{tabular}} (idle);

		\path (execT1) edge node{\begin{tabular}{@{} c @{\ } c@{} }
		 & $\styleact{actT5}$\\
		\end{tabular}} (execT1waitT5);

		\path (execT1) edge[bend left] node{\begin{tabular}{@{} c @{\ } c@{} }
		& $ \styleclock{xactT1} > \styleparam{T1Period}$\\
		 & $\styleact{DeadlineMiss}$\\
		\end{tabular}} (DeadlineMissed);

		\path (execT1waitT5) edge[bend left] node[right]{\begin{tabular}{@{} c @{\ } c@{} }
		& $ \styleclock{xexecT1} \geq \styleparam{T1BCET}$\\
		 & $\styleact{finT1}$\\
		 & $\styleclock{xactT1}:=0$\\
		\end{tabular}} (execT5);

		\path (execT1waitT5) edge node{\begin{tabular}{@{} c @{\ } c@{} }
		& $ \styleclock{xactT5} > \styleparam{T5Period} $\\
		& $ \lor \styleclock{xactT1} > \styleparam{T1Period} $\\
		 & $\styleact{DeadlineMiss}$\\
		\end{tabular}} (DeadlineMissed);

		\path (execT5) edge[bend left] node{\begin{tabular}{@{} c @{\ } c@{} }
		& $ \styleclock{xexecT5} \geq \styleparam{T5BCET}$\\
		 & $\styleact{finT5}$\\
		 & $\styleclock{xactT5}:=0$\\
		\end{tabular}} (idle);

		\path (execT5) edge[bend left] node[right]{\begin{tabular}{@{} c @{\ } c@{} }
		 & $\styleact{actT1}$\\
		\end{tabular}} (execT1waitT5);

		\path (execT5) edge[bend right] node[below right]{\begin{tabular}{@{} c @{\ } c@{} }
		& $ \styleclock{xactT5} > \styleparam{T5Period}$\\
		 & $\styleact{DeadlineMiss}$\\
		\end{tabular}} (DeadlineMissed);
 
	\end{tikzpicture}
		}

	\caption{Translating the scheduler CPU1 with a preemptive \FPS{} scheduling policy}
	\label{figure:translation:scheduler}
\end{figure*}
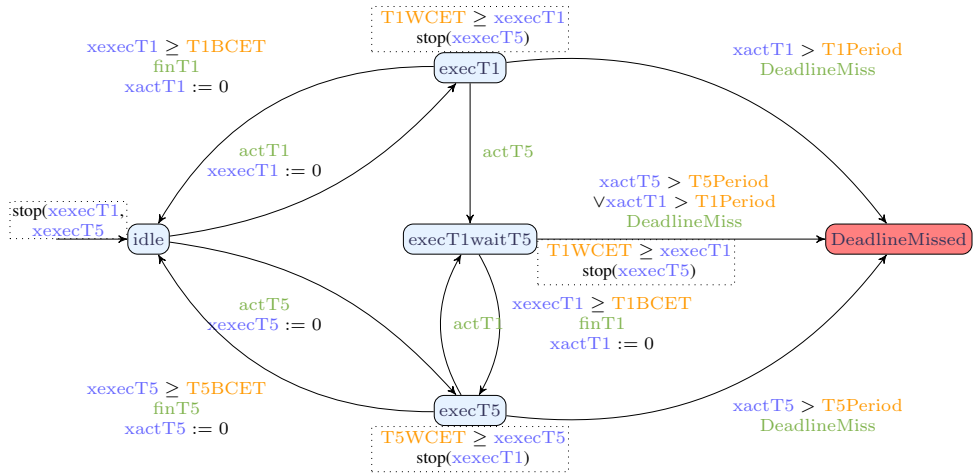

Our encoding of \FPS{} is a formalization of the \emph{ad-hoc} scheme of~\cite{SSLAF13}.
\cref{figure:translation:scheduler} shows the translation of CPU1 from \cref{figure:example:Time4sys}, assuming its policy is preemptive \FPS{}, with the priority of Task~1 being higher than that of Task~5.
We use stopwatches, and the stopped clocks in a location are given in the dashed rectangle together with the invariant.
We assume task $T_1$ has higher priority over~$T_5$.
The processor starts by being \styleloc{idle}, waiting for a task activation.
As soon as a task is activated (\eg{} action \styleact{actT5}), it moves to one of the locations where the corresponding task is running (\styleloc{execT5}).
If it receives another activation request (\styleact{actT1}), it moves to the location corresponding to the higher priority task running (\styleact{execT1waitT5}), where $T_1$ is executed and~$T_5$ is waiting to be executed.
The fact that $T_5$ does not execute anymore is modeled by the stopping of the clock $\styleclock{xexecT5}$ corresponding to task~$T_5$.
Moreover, while a task executes, the scheduler automaton checks if it misses its deadline (\eg{} guard $\styleclock{xactT5} > \styleparam{T5Period}$, where $\styleparam{T5Period}$ is $T_5$'s deadline since we assumed deadlines to be equal to periods).
In the case of a deadline miss, the processor PTA moves to a special failure location (``deadline missed'') and stops any further computation.

The model of a non-preemptive processor is very similar to the model of preemptive processor: the central location in
\cref{figure:translation:scheduler} which accounts for the fact that $T_5$ is stopped when $T_1$ is activated, in the non-preemptive case must not stop $T_5$, but simply remember that $T_1$ has been released, so that we can move to the top state when $T_5$ completes its instance.

\paragraph*{Encoding \TDMA{}}
We also encoded the ``Time-division multiple access'' (\TDMA{}) scheduling policy.
This scheduling policy is fairly simple, as there are no priorities: the scheduler simply passes a ``token'' to each of its tasks in a circular manner.
If an instance of this task is activated, then this task executed for a predefined amount of time (generally small).
If no instance of this task is activated, then the processor remains idle during the predefined amount of time, before moving to the next task.

\paragraph*{Other scheduling policies}
Other scheduling policies are discussed in \cref{section:conclusion}.

\subsection{Handling uncertainty}

A major interest of our transformation is its ability to handle \emph{uncertainty}.
Indeed, all timing constants (periods, deadlines, offsets…)\ can be possibly kept parametric, or can be constrained to belong to a predefined interval.
For example, to model the period of a task~$T_1$ equal to 100\,ms known with a precision of 1\,\%, it suffices to declare the parameter $\styleparam{T1Period} \in [99, 101]$, which can be simulated by a small PTA gadget---or is natively offered by the input syntax of \imitator{}.

Then, the system is schedulable for all valuations such that the system reaches none of the \styleloc{DeadlineMissed} locations defined in \cref{ss:scheduling}.
This can be computed by the \EFsynth{} algorithm implemented in \imitator{}.

\subsection{Implementation}\label{ss:implementation}

We implemented our translation in a Java program of 2,900~lines of code called \JawherTool{}~\cite{Jawher19}.
The program takes as input a \TimeForSys{} model exported from the interface (that comes in the form of an XML file), and outputs a PTA model in the \imitator{} input syntax.
We support all the features mentioned in \cref{section:transformation}.

\reviewer{1}{For the research paper category, thorough analysis of the technique together with pros
and cons are needed (but not given by the paper in the current form): questions like:
1) does the technique scale? 2) what is the impact on the scheduler chosen on the size
of the (translated) model? 3) What is the complexity of the translation in terms of
some input-to-output model size? 4) Are there alternatives in the translation process
and what would be their impact? And finally, one would like to see some figures on
how long does the verification or identification of some admissible values for parameters
takes for some benchmark-like models/problems?}

\section{Experiments}\label{section:experiments}

As a proof of concept, we apply our transformation to the example in \cref{figure:example:Time4sys}.
Task~1 (resp.~5) has an activation period of 10 (resp.~20) and an offset of~5 (resp.~0).
Task~6 is sporadic with a minimum interarrival time of~20.
The best and worst case computation times for the tasks are given in bracket; for example, the computation time of each instance of task~1 is non-deterministically chosen in $[4,5]$.
Recall that all three processors use \FPS{}.
The priorities are chosen as follows:
	for CPU1, $T1 > T5$;
	for CPU2, $T2 > T4 > T7$;
	for CPU2, $T6 > T3$.

Our translation yields a model made of 14 synchronization actions, 14 clock variables, and (up to) 28 parameters, that consist of all periods, deadlines, offsets and best and worst case execution times.
\imitator{} allows to keep none, some or all of the parameters unconstrained to allow for various levels of parameterization.

Experiments were conducted using \imitator{} 2.10.4 ``Butter Jellyfish'' %
on a Dell XPS\,13 9365 i7 (2017) running Linux Mint 18.2 64\,bits with 8\,GiB memory.\footnote{%
	Experimental results with models, logs, results and graphics can be found at \url{https://www.imitator.fr/static/TASE19}.
}

\paragraph{Non-parametric schedulability analysis}
First, we valuate all parameters (therefore the model is entirely non-parametric), and we assume that all deadlines are equal to period for periodic tasks (we are not interested in deadline violations for other tasks, therefore their deadline can be set to an arbitrarily large value).
An analysis with \imitator{} (computation time: 0.2\,s) certifies that this model is schedulable, \ie{} no deadline miss occurs.

\begin{figure}[tb]
	\centering
	\footnotesize
	\begin{subfigure}[b]{.48\columnwidth}
		
		\centering
		\includegraphics[width=.8\textwidth]{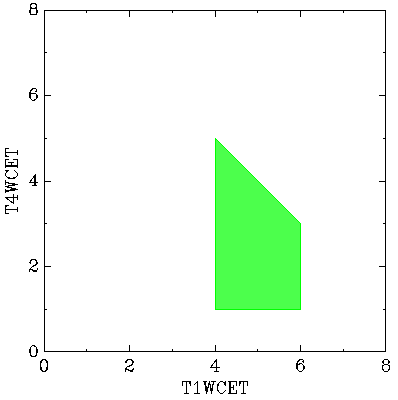}

		\caption{WCET 1 and~4}
		\label{figure:cartography:WCET1-WCET4}
	\end{subfigure}
	\begin{subfigure}[b]{.48\columnwidth}
		
		\centering
		\includegraphics[width=.8\textwidth]{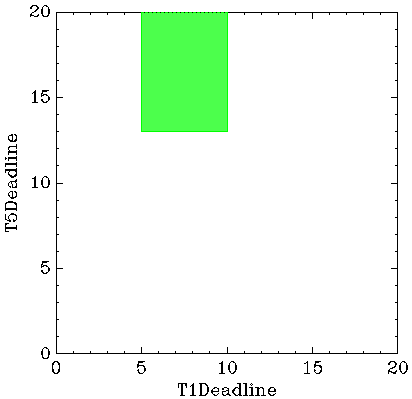}

		\caption{Deadlines 1 and~5}
		\label{figure:cartography:Deadlines1-5}
	\end{subfigure}
	
	\caption{Some 2-D graphical representations}
	\label{figure:cartographies}
\end{figure}

\paragraph{Parametric WCETs}
Then, we leave some parameters unconstrained, so as to not only verify the schedulability but also to \emph{synthesize} parameter values for which the system is schedulable.
Leaving \styleparam{T1WCET} and \styleparam{T4WCET} unconstrained yields the following constraint (time: 1.5s):\footnote{%
	While \imitator{} may sometimes output incomplete constraints, all constraints given in this manuscript are certified exact (sound and complete) by the tool.
}

$
4 \leq \styleparam{T1WCET} \leq 6 \land \styleparam{T4WCET} \geq 1
\\
\land \styleparam{T1WCET} + \styleparam{T4WCET} < 9
$

A graphical representation (output by \imitator{}) is given in \cref{figure:cartography:WCET1-WCET4}.

Then, we run a 4-dimensional analysis to infer values for \styleparam{T1WCET}, \styleparam{T4WCET}, \styleparam{T5WCET} and \styleparam{T7WCET} (time: 11.5s);
the result is given in \cref{figure:constraint:4WCET}.

\begin{figure*}[tb]
	\centering
	\footnotesize
	
	\begin{subfigure}[b]{.32\textwidth}
		$
		\styleparam{T5WCET} \geq 6
		\land \styleparam{T4WCET} \geq 1\\
		\land 20 \geq 2*\styleparam{T1WCET} + \styleparam{T5WCET}\\
		\land 9 > \styleparam{T1WCET} + \styleparam{T4WCET}\\
		\land \styleparam{T1WCET} \geq 4
		\land \styleparam{T7WCET} \geq 10\\
		\lor\\
		\styleparam{T5WCET} \geq 6 
		\land \styleparam{T7WCET} \geq 10\\
		\land 16 > 2*\styleparam{T4WCET} + \styleparam{T7WCET}\\
		\land \styleparam{T1WCET} + \styleparam{T4WCET} \geq 9\\
		\land 20 \geq 2*\styleparam{T1WCET} + \styleparam{T5WCET}
		$
		\caption{4-dimensional parametric WCETs}
		\label{figure:constraint:4WCET}
	\end{subfigure}
	\begin{subfigure}[b]{.34\textwidth}
		\footnotesize
	$
	\styleparam{T5Deadline} \geq \styleparam{T1WCET} + \styleparam{T5WCET}\\
	\land \styleparam{T1WCET} \geq 4\\
	\land \styleparam{T1Deadline} \geq \styleparam{T1WCET}\\
	\land 15 \geq \styleparam{T1WCET} + \styleparam{T5WCET}\\
	\land \styleparam{T5Deadline} + 5 \geq 2*\styleparam{T1WCET} + \styleparam{T5WCET}\\
	\land 20 \geq \styleparam{T5Deadline}\\
	\land 10 \geq \styleparam{T1Deadline}\\
	\land \styleparam{T5WCET} \geq 6\\
	\land 20 \geq 2*\styleparam{T1WCET} + \styleparam{T5WCET}\\
	\lor \\
	10 \geq \styleparam{T1Deadline}
	\land \styleparam{T1Deadline} \geq \styleparam{T1WCET}\\
	\land \styleparam{T1WCET} + \styleparam{T5WCET} > 15\\
	\land \styleparam{T1WCET} \geq 4\\
	\land \styleparam{T5Deadline} \geq 2*\styleparam{T1WCET} + \styleparam{T5WCET}\\
	\land 20 \geq \styleparam{T5Deadline}
	$
	\caption{WCETs and deadlines}
	\label{figure:constraint:2WCET-2DEAD}
	\end{subfigure}
	\begin{subfigure}[b]{.32\textwidth}
		$
		13 > \styleparam{T1WCET} + \styleparam{T1Offset}\\
		\land 6 > \styleparam{T1WCET} \geq 4\\
		\land 10 \geq \styleparam{Deadline}\\
		\land \styleparam{Deadline} \geq \styleparam{T1WCET}\\
		\land \styleparam{T1Offset} > 1\\
		\lor \\
		    7 > \styleparam{T1Offset} > 3 \\
		\land 10 \geq \styleparam{Deadline} \geq 6 \\
		\land \styleparam{T1WCET} = 6
		$

		\caption{Task~1: offset, WCET, deadline}
		\label{figure:constraint:3DTask1}
	\end{subfigure}

	\caption{Results}
	\label{figure:constraints}
\end{figure*}

\paragraph{Parametric deadlines}
We then parameterize deadlines of tasks~1 and~5.
The result (computed in 0.63\,s) is\LongVersion{ given below}:
\LongVersion{

}%
$
\styleparam{T1Deadline} \in [5, 13] 
\land
 \styleparam{T5Deadline} \in [10, 20]
$

A graphical representation is given in \cref{figure:cartography:Deadlines1-5}.

\paragraph{Parametric offset}
We also study the influence of the offset of Task~1 on the schedulability.
An analysis with one parameter \styleparam{OffsetT1} gives (time: 2.0\,s):
$\styleparam{OffsetT1} \in (1, 8)$.

It may be surprising that the system is non-schedulable for a zero-offset: this comes from the fact that, if Task~1 is activated too soon, Task~2 (that is activated upon completion of Task~1) will preempt CPU2, preventing Task~7 to complete, notably in the worst case when Task~6 is activated at the highest possible period allowed by its sporadic nature and Task~7 lasts for its longest duration (15\,time units).

\paragraph{Multidimensional parametric analyses}
We finally relate the various variables of the system by performing parametric analyses in different dimensions.
We first relate deadlines and WCETs of tasks~1 and~5.
The analysis (time: 8.9\,s) gives the result in \cref{figure:constraint:2WCET-2DEAD}.

Finally, we relate three values of task~1, \ie{} its offset, its WCET and its deadline.
We give the result (time: 40.6\,s) in \cref{figure:constraint:3DTask1}.

\section{Perspectives}\label{section:conclusion}

\todo{référence à l'outil de Jawher}

In this work, we proposed a first translation of an industrial formalism to model real-time systems into a parametric timed formalism, namely parametric timed automata.
We discuss below short-term perspectives.

\paragraph*{Lifting assumptions}

\LongVersion{%
	We did not support the full syntax of \TimeForSys{}, as we used some assumptions defined in \cref{ss:assumptions}.
}

We assumed that all periodic tasks must have a deadline equal to the period.
While deadlines less than or equal to the period is very straightforward (and is part of our implementation), deadlines larger than periods may render the translation more elaborate.
Indeed, our encoding of \cref{ss:taskchains} needs to be deeply modified, as we need to encode more than one instance of a task at a given time.
An option is to split further the task chains, and add some additional integer-valued variables, but this solution will come at the cost of a less efficient analysis.

Multiple dependencies remain to be supported: on the one hand, a task activating several tasks (\eg{} in \cref{figure:GUI}, \texttt{Tracking\_control2} upon completion activates both \texttt{Camera\_control} and \texttt{Tracking\_control3}) is natural and should be supported without difficulties by action synchronization\LongVersion{ in parametric timed automata};
on the other hand, the semantics of several tasks activating a single task looks more dubious, and we are not even sure to want to support this syntax.

Jitters can easily be supported: however, they slightly complicate the model of task activation patterns, and are therefore discarded both in this work and temporarily in our implementation (with the goal to introduce them soon).

After lifting these assumptions, we note that we support the vast majority of the \TimeForSys{} syntax, with the exception of multiple input task activation---which, again, may seem dubious as its semantics is very unclear.

\paragraph*{Alternative scheduler models}
A natural future work will be to propose optimizations or variants in the translation, notably of the scheduler model, and test their practical efficiency on a set of benchmarks, such as the \imitator{} benchmarks library~\cite{Andre18FTSCS}.

\LongVersion{
Implementing other scheduling policies, such as earliest deadline first (\EDF{}) or shortest job first (\SJF{}), should be straightforward.
Round Robin resembles \TDMA{} in the sense that the scheduler moves a ``token'' to its tasks; however, the main difference is that, whenever a task has no activated instance, the processor immediately moves to the next one.
This is not difficult as such, but this requires more locations \LongVersion{in the PTA model }than for \TDMA{}.
}

\paragraph*{Translation to \uppaal{}}

A translation into the state-of-the-art \uppaal{} model-checker~\cite{LPY97} is on our agenda.
However, a translation to \uppaal{} would lose the ability to use unknown or uncertain constants; in addition, \uppaal{} does not fully support stopwatches, needed to model preemption.

\paragraph*{Execution traces}
Finally, in case of a deadline miss, we aim at displaying the faulty model trace leading to the deadline miss back to the \TimeForSys{} model.
Note that \TimeForSys{} also features a metamodel for such traces.

\todo{R1: ``Some solutions such as priority ceiling and priority inheritance can be used to avoid these deadlocks.''}

\section*{Acknowledgements}

\todo{logo soutien région île de France}

The author is grateful to
Romain Soulat (Thales Research and Technology, Palaiseau) for interactions concerning \TimeForSys{},
to Jawher Jerray  and Sahar Mhiri for their help on the implementation of the translation of \TimeForSys{} into \imitator{},
and to the reviewers for helpful suggestions.

\ifdefined\VersionLong
	\newcommand{\CCIS}{Communications in Computer and Information Science}
	\newcommand{\ENTCS}{Electronic Notes in Theoretical Computer Science}
	\newcommand{\FI}{Fundamenta Informormaticae}
	\newcommand{\FMSD}{Formal Methods in System Design}
	\newcommand{\IJFCS}{International Journal of Foundations of Computer Science}
	\newcommand{\IJSSE}{International Journal of Secure Software Engineering}
	\newcommand{\IPL}{Information Processing Letters}
	\newcommand{\JLAP}{Journal of Logic and Algebraic Programming}
	\newcommand{\JLC}{Journal of Logic and Computation}
	\newcommand{\LMCS}{Logical Methods in Computer Science}
	\newcommand{\LNCS}{Lecture Notes in Computer Science}
	\newcommand{\RESS}{Reliability Engineering \& System Safety}
	\newcommand{\STTT}{International Journal on Software Tools for Technology Transfer}
	\newcommand{\TCS}{Theoretical Computer Science}
	\newcommand{\ToPNoC}{Transactions on Petri Nets and Other Models of Concurrency}
	\newcommand{\TSE}{IEEE Transactions on Software Engineering}
	\renewcommand*{\bibfont}{\small}
	\printbibliography[title={References}]
\else
	\bibliographystyle{IEEEtran} %
	\newcommand{\CCIS}{CCIS}
	\newcommand{\ENTCS}{ENTCS}
	\newcommand{\FI}{FI}
	\newcommand{\FMSD}{FMSD}
	\newcommand{\IJFCS}{IJFCS}
	\newcommand{\IJSSE}{IJSSE}
	\newcommand{\IPL}{IPL}
	\newcommand{\JLAP}{JLAP}
	\newcommand{\JLC}{JLC}
	\newcommand{\LMCS}{LMCS}
	\newcommand{\LNCS}{LNCS}
	\newcommand{\RESS}{RESS}
	\newcommand{\STTT}{STTT}
	\newcommand{\TCS}{TCS}
	\newcommand{\ToPNoC}{ToPNoC}
	\newcommand{\TSE}{TSE}
	\bibliography{Time4sys}
\fi
\LongVersion{
\appendix
\subsection{Examples with deadline misses}
\subsubsection{Modifying the deadline}\label{appendix:miss1}

Consider again the real-time system in \cref{figure:example:Time4sys}.
Now observe that, if the deadline of Task~5 was 11 (instead of~20), then at $t = 11$, a deadline miss would occur for Task~5, as the execution of its instance is not completed by its deadline.
This situation is depicted in \cref{figure:chronogramme:miss1}: the deadline is depicted using a down arrow, and the part of the execution of the instance of Task~5 violating the deadline is highlighted using a red box.

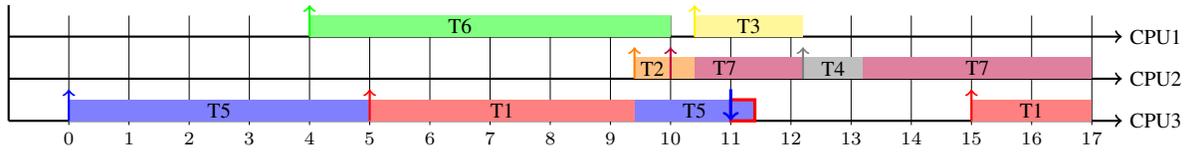
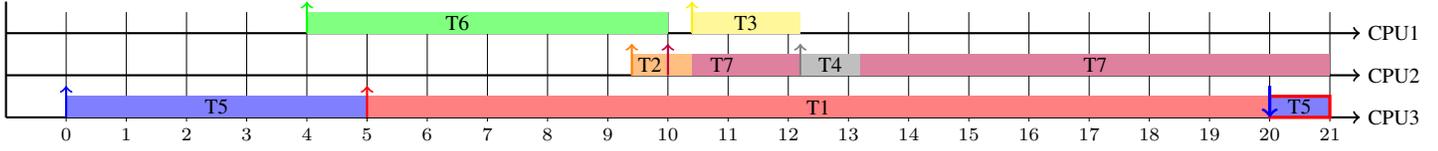
\begin{figure*}
	\begin{subfigure}[b]{.99\textwidth}
		
		\centering
		\begin{tikzpicture}[scale=.8, xscale=1, yscale=.7]
			\footnotesize

			\draw[thick] (-1, 0) --++ (0, 2.75);
			\draw[thick,->] (-1, 2) -- (17.5,2) node[right] {CPU1};
			\draw[thick,->] (-1, 1) -- (17.5,1) node[right] {CPU2};
			\draw[thick,->] (-1, 0) -- (17.5,0) node[right] {CPU3};

			\foreach \x in {0, 1, ..., 17} %
				\draw[very thin] (\x, 2.5) -- (\x, -.1) node [below] {\scriptsize $\x$};

			\draw[draw=none,fill=blue!50] (0, 0) rectangle (5, .5);
			\node at (2.5, .25) {T5};
			
			\draw[draw=none,fill=red!50] (5, 0) rectangle (9.4, .5);
			\node at (5/2 + 9.4/2, .25) {T1};
			
			\draw[draw=none,fill=blue!50] (9.4, 0) rectangle (11.4, .5);
			\node at (10.4, .25) {T5};
			\draw[violation] (11.0, 0) rectangle (11.4, .5);
			
			\draw[draw=none,fill=red!50] (15, 0) rectangle (17, .5);
			\node at (16, .25) {T1};

			\draw[draw=none,fill=orange!50] (9.4, 1) rectangle (10.4, 1.5);
			\node at (9.7, 1.25) {T2};
			
			\draw[draw=none,fill=purple!50] (10.4, 1) rectangle (12.2, 1.5);
			\node at (10.9, 1.25) {T7};
			
			\draw[draw=none,fill=gray!50] (12.2, 1) rectangle (13.2, 1.5);
			\node at (12.7, 1.25) {T4};
			
			\draw[draw=none,fill=purple!50] (13.2, 1) rectangle (17, 1.5);
			\node at (13.2/2 + 17/2, 1.25) {T7};

			\draw[draw=none,fill=green!50] (4, 2) rectangle (10, 2.5);
			\node at (6.5, 2.25) {T6};
			
			\draw[draw=none,fill=yellow!50] (10.4, 2) rectangle (12.2, 2.5);
			\node at (10.4/2 + 12.2/2, 2.25) {T3};

			\draw[activation, draw=blue] (0, 0) --++ (0,.75);
			\draw[activation, draw=red] (5, 0) --++ (0,.75);
			\draw[activation, draw=red] (15, 0) --++ (0,.75);
			
			\draw[activation, draw=orange] (9.4, 1) --++ (0,.75);
			\draw[activation, draw=purple] (10, 1) --++ (0,.75);
			\draw[activation, draw=gray] (12.2, 1) --++ (0,.75);
			
			\draw[activation, draw=green] (4, 2) --++ (0,.75);
			\draw[activation, draw=yellow] (10.4, 2) --++ (0,.75);

			\draw[deadline, draw=blue] (11, 0.75) --++ (0,-.75);

		\end{tikzpicture}

		\caption{Deadline miss for \cref{figure:example:Time4sys} if $\styleparam{T5Deadline} = 11$}
		\label{figure:chronogramme:miss1}
		
	\end{subfigure}

	\begin{subfigure}[b]{.99\textwidth}
		\centering
		\begin{tikzpicture}[scale=.8, xscale=1, yscale=.7]
			\footnotesize

			\draw[thick] (-1, 0) --++ (0, 2.75);
			\draw[thick,->] (-1, 2) -- (21.5,2) node[right] {CPU1};
			\draw[thick,->] (-1, 1) -- (21.5,1) node[right] {CPU2};
			\draw[thick,->] (-1, 0) -- (21.5,0) node[right] {CPU3};

			\foreach \x in {0, 1, ..., 21} %
				\draw[very thin] (\x, 2.5) -- (\x, -.1) node [below] {\scriptsize $\x$};

			\draw[draw=none,fill=blue!50] (0, 0) rectangle (5, .5);
			\node at (2.5, .25) {T5};
			
			\draw[draw=none,fill=red!50] (5, 0) rectangle (20, .5);
			\node at (12.5, .25) {T1};
			
			\draw[draw=none,fill=blue!50] (20, 0) rectangle (21, .5);
			\node at (20.5, .25) {T5};
			\draw[violation] (20.0, 0) rectangle (21, .5);

			\draw[draw=none,fill=orange!50] (9.4, 1) rectangle (10.4, 1.5);
			\node at (9.7, 1.25) {T2};
			
			\draw[draw=none,fill=purple!50] (10.4, 1) rectangle (12.2, 1.5);
			\node at (10.9, 1.25) {T7};
			
			\draw[draw=none,fill=gray!50] (12.2, 1) rectangle (13.2, 1.5);
			\node at (12.7, 1.25) {T4};
			
			\draw[draw=none,fill=purple!50] (13.2, 1) rectangle (21, 1.5);
			\node at (13.2/2 + 21/2, 1.25) {T7};

			\draw[draw=none,fill=green!50] (4, 2) rectangle (10, 2.5);
			\node at (6.5, 2.25) {T6};
			
			\draw[draw=none,fill=yellow!50] (10.4, 2) rectangle (12.2, 2.5);
			\node at (10.4/2 + 12.2/2, 2.25) {T3};

			\draw[activation, draw=blue] (0, 0) --++ (0,.75);
			\draw[activation, draw=red] (5, 0) --++ (0,.75);
			
			\draw[activation, draw=orange] (9.4, 1) --++ (0,.75);
			\draw[activation, draw=purple] (10, 1) --++ (0,.75);
			\draw[activation, draw=gray] (12.2, 1) --++ (0,.75);
			
			\draw[activation, draw=green] (4, 2) --++ (0,.75);
			\draw[activation, draw=yellow] (10.4, 2) --++ (0,.75);

			\draw[deadline, draw=blue] (20, 0.75) --++ (0,-.75);

		\end{tikzpicture}

		\caption{Deadline miss for \cref{figure:example:Time4sys} if $\styleparam{T1WCET} = 15$}
		\label{figure:chronogramme:miss2}
	\end{subfigure}
	
	\caption{Two examples of deadline misses}

\end{figure*}
\subsubsection{Modifying the WCET}\label{appendix:miss2}

Alternatively, if the deadline of Task~5 was 20 as in the original definition but the worst case execution time of Task~1 was 15 (with an increased period and deadline for Task~1 of, say, 20), then at $t=20$, Task~5 has still not completed and, again, a deadline miss occurs.
This situation is depicted in \cref{figure:chronogramme:miss2}.

}

\end{document}